\documentclass[journal]{IEEEtran}
\ifCLASSINFOpdf
\else
\fi

\usepackage{graphicx} 
\usepackage{subfigure}
\usepackage{epsfig} 
\usepackage{balance}
\usepackage{yhmath} 

\usepackage{times} 
\usepackage{amsmath} 
\usepackage{amssymb}  
\usepackage{algorithm}
\usepackage{algorithmic}

\newtheorem{lemma}{Lemma}
\newtheorem{theorem}{Theorem}

\newtheorem{remark}{Remark}
\newtheorem{myDef}{Definition}

\begin{document}
%
\title{Set Covering-based Approximation Algorithm for Delay Constrained Relay Node Placement in Wireless Sensor Networks}
%
%
%

\author{Chaofan~Ma,
        Wei~Liang,~\IEEEmembership{Member,~IEEE,}
        and~Meng~Zheng,~\IEEEmembership{Member,~IEEE,}
\thanks{This work was supported by the Natural Science Foundation of China (61174026, 61233007, and 61304263) and Cross-disciplinary Collaborative Teams Program for Science, Technology and Innovation of Chinese Academy of Sciences (Network and System Technologies for Safety Monitoring and Information Interacting in Smart Grid).}
\thanks{Chaofan Ma, Wei Liang, and Meng Zheng are with the Key Laboratory of Networked Control Systems, Chinese Academy of Sciences, Shenyang 110016, China. E-mail: \{machaofan, weiliang, zhengmeng\_6\}@sia.cn.}
\thanks{Chaofan Ma is also with the University of Chinese Academy of Sciences, Beijing 100049, China.}}

\maketitle

\begin{abstract}
The Delay Constrained Relay Node Placement (DCRNP) problem in Wireless Sensor Networks (WSNs) aims to deploy minimum
relay nodes such that for each sensor node there is a path connecting this
sensor node to the sink without violating delay constraint. As WSNs are gradually employed in time-critical applications, the
importance of the DCRNP problem becomes noticeable. For the NP-hard nature of DCRNP problem, an approximation
algorithm-Set-Covering-based Relay Node Placement (SCA) is proposed to solve the
DCRNP problem in this paper.
The proposed SCA algorithm deploys relay nodes iteratively from sink to the given
sensor nodes in hops, i.e., in the $k$th iteration SCA deploys relay nodes at the
locations that are $k$ hops apart from the sink.
Specifically, in each iteration, SCA first finds the candidate deployment locations
located within 1 hop to the relay nodes and sensor nodes, which have already
been connected to the sink. Then, a subset of these candidate deployment locations, which
can guarantee the existence of paths connecting unconnected sensor nodes to the sink within delay constraint,
is selected to deploy relay nodes based on the set covering method. As the iteration of SCA algorithm,
the sensor nodes are gradually connected to the sink with satisfying delay constraint.

The elaborated analysis of the approximation ratio of SCA algorithm is given out, and we also
prove that the SCA is a polynomial time algorithm through rigorous time complexity
analysis. To evaluate the performance of the proposed SCA algorithm, extensive
simulations are implemented, and the simulation results show that the SCA algorithm
can significantly save the deployed relay nodes comparing to the existing algorithms, i.e., at most
$31.48\%$ deployed relay nodes can be saved due to SCA algorithm.
\end{abstract}

\begin{IEEEkeywords}
Wireless sensor networks, set covering, delay constrained relay node placement,
approximation algorithm.
\end{IEEEkeywords}

%
\IEEEpeerreviewmaketitle

\section{Introduction}
%
%
%
%
\IEEEPARstart{W}{ireless}
Sensor Networks (WSNs) attract considerable attention in recent years for their immense
potential in the applications of environment monitoring, battlefield reconnaissance and industrial
automation \cite{Akyildiz02}-\cite{Estrin99}. WSNs comprise spatially distributed Sensor Nodes
(SNs), which sense certain circumstance information, and one or several sinks to collect the information
sensed by SNs. Typically, SNs are cheap and powered by batteries. This results in that the
power and communication radii of a SN is limited. Thus, to prolong the network lifetime and improve the
scalability, Relay Nodes (RN) are introduced into WSNs.
RNs are equipped with plenty power and large communication radii, however,
the RNs are also costly \cite{Yick08}-\cite{Bao09}. Hence, the Relay Node Placement (RNP) problem
try to build connected WSNs by deploying a minimum number of RNs subject to various constraints,
such as network lifetime, throughput and delay.

Recently, the importance of Delay Constrained RNP (DCRNP)
problem is highlighted by the employment of WSNs in the time-critical applications, e.g., the factory automation and patent monitoring
\cite{Su09}-\cite{Chen12}.
In the factory environment, the data sensed by SNs is typically time-sensitive, such as alarm notification and information for feedback control,
and thus the importance of receiving the data at the sink in a timely manner is noticeable \cite{Gungor09}-\cite{Kumar14}.
It has been reported in \cite{Bhattacharya10} that DCRNP problem is NP-hard, which implies that no polynomial time algorithm exists for this problem as long as $P\neq NP$.

Although RNP problem has been extensively studied, the literature about DCRNP problem is rare. Bhattacharya and Kumar
\cite{Bhattacharya10}-\cite{Bhattacharya14} first proved that DCRNP problem is NP-hard, and then proposed a polynomial time approximation algorithm to solve this problem. Nigam and Agarwal \cite{Nigam14} designed an algorithm to optimally solve
DCRNP problem without guaranteeing a polynomial time complexity. As in \cite{Bhattacharya10}-\cite{Nigam14}, in this paper, RNs can only be placed at a subset of the predetermined locations, which are called the Candidate Deployment Locations (CDLs).

In order to devise a polynomial time algorithm with better performance,
a Set-Covering-based Approximation (SCA) algorithm is proposed
in this paper to yield a desirable solution for DCRNP problem. The proposed SCA is an iterative
algorithm and deploys RNs in hops from the sink to SNs. In order to maintain network connectivity, in each iteration, the RNs that are deployed based on set covering method have 1-hop neighbor SNs or RNs, which have been connected to the sink within delay constraint. As SCA algorithm is iteratively executed, SNs are gradually connected to the sink with fulfilling delay constraint.
To evaluate the proposed SCA algorithm, extensive simulations are implemented, and the
simulation results show that the SCA algorithm outperforms the existing algorithms outstandingly.

The major contributions of this paper are listed as follows:
\begin{itemize}
  \item An approach is designed to represent each CDL or SN by the unconnected SNs, which
  can be connected to the sink by the paths that fulfil delay constraint and pass through this CDL or SN.
  \item An approximation algorithm-SCA is proposed for DCRNP problem. In each iteration, SCA selects a subset of
  the CDLs, which are located within 1-hop from the RNs and SNs that have been connected to the sink, to deploy
  RNs by solving a set covering problem that are modeled by the approach mentioned above. Due to this manner,
  SCA can avoid the limitation suffering by the algorithm presented in \cite{Bhattacharya10}-\cite{Bhattacharya14}.
  \item The approximation ratio of SCA algorithm is explicitly analyzed, and the time complexity
  of SCA algorithm is also detailed in this paper.
\end{itemize}

This paper is organized as follows. Section II reviews the related works. The problem formulation is
given out in Section III. The preliminaries and heuristic are given out in Section IV. Section V
describes our proposed algorithm for the DCRNP problem. Section VI shows an explicit analysis of approximation ratio
and time complexity of the proposed heuristic algorithm. The efficiency of the proposed algorithm is validated through
extensive simulations in Section VII. Finally, Section VIII concludes the whole paper.

\section{Related Work}
The related literatures are divided into two
categories, i.e., the literatures about the DCRNP problem and the RNP problem without delay constraint.
\subsection{RNP Problem without Delay Constraint}
Lin and Xue \cite{Lin99} studied the RNP problem and formulated it as the as the Steiner minimum tree with minimum number
of Steiner points and bounded edge length problem (SMT-MSP). Then, they proved the SMT-MSP to be NP-Complete
and proposed a 5-approximation algorithm. Chen et al. \cite{Chen00} demonstrated that the algorithm proposed in \cite{Lin99}
is actual a 4-approximation algorithm, and proposed a 3-approximation algorithm for the RNP problem. Cheng et al. \cite{Cheng07} presented a 3-approximation algorithm and a 2.5-approximation algorithm based on so-called 3-star structure. Lloyd and Xue \cite{Lloyd07} presented a 7-approximation
algorithm for the single-tired network RNP problem and a $(5+\epsilon)$-approximation algorithm for the two-tired
network RNP problem. Srinivas et al. \cite{Srinivas09} studied the problem of constructing and maintaining the wireless
backbone network for the WSNs, and they also considered mobile wireless backbone network.

Misra et al. \cite{Misra08}-\cite{Misra10} studied the constrained RNP problem in the single-tiered network and proposed a
polynomial time O(1)-approximation algorithm. Yang et al. \cite{Yang12} first studied the constrained RNP problem
in the two-tiered network and proposed an algorithm with O(1)-approximation ratio for 1-connected single cover problem. Then
they proposed two algorithms with O(1)-approximation ratio and O(ln$n$)-approximation ratio under different settings for 2-connected double cover problem.

\subsection{DCRNP Problem}
The DCRNP problem has been studied in \cite{Bhattacharya10}-\cite{Nigam14}. Bhattacharya and Kumar
\cite{Bhattacharya10}-\cite{Bhattacharya14} first proved that the DCRNP problem is NP-hard, and a
shorted path tree based algorithm
is proposed. This algorithm preliminarily forms a shortest path tree to
connect each SN to the sink, then it saves the deployed RNs by gradually removing the
RNs on the shortest path tree. This brings in a limitation that the deployed RNs can only be
those contained in the originally yielded shortest path tree, and the worst case of this algorithm happens
when all the RNs of the optimal solution have been missed by the shortest path tree.
Sitanayah et al. \cite{Sitanayah12} studied the fault-tolerant RNP problem under the length constraint, and proposed a heuristic
to solve this problem. However, no time complexity analysis and performance guarantee were given.
Nigam and Agarwal
\cite{Nigam14} formulated the DCRNP problem as a linear programming problem, and proposed a branch-and-cut
algorithm to optimally solve the DCRNP problem. However, the proposed algorithm can only solve a special
case of the DCRNP problem (each of the source node cannot have a singleton node cut), and the time complexity
of this algorithm grows exponentially, which indicates it cannot be applied to the large scale problem.

Some researches on the related scopes are also listed in follows. Voss \cite{Voss99} used the hop count to represent
the network delay, and studied the hop constrained Steiner tree problem. A local search algorithm based on
tabu search was proposed, but this paper did not analyze the approximation ratio and the time complexity.
Costa et al. \cite{Costa08} studied the Steiner tree problem with revenue, budget and hop constraints. A
tabu search based heuristic was proposed to solve this problem, but the proposed
heuristic is not a polynomial time algorithm. Gouveia et al. \cite{Gouveia08} studied the distance constrained
minimum spanning tree problem, and three approaches were proposed based on the column generation scheme and
the Lagrangian relaxation. However, no time complexity analysis is provided in the paper.

\section{Problem Formulation}
The network delay is typically composed of the processing delay, the queuing delay, the transmission delay and the
propagation delay. In the WSNs, the distance between two different SNs is normally a few tens or hundreds of meters.
Therefore, normally
compared to the other three kinds of delay, the transmission delay can be ignored. Since the rest of delays
are at proportional to the hop count, in this paper, the network delay is represented by the hop count. Therefore,
DCRNP problem is reduced to a hop constrained RNP problem in the rest of this paper.

Given a set of SNs $S=\{s_1,s_2,...,s_n\}$, a set of Candidate Deployment Locations (CDLs) $C=\{c_1,c_2,...,c_m\}$ for
deploying RNs and
a sink $K$, we can build an undirected graph $G=\{V,E\}$, where
$V=S\bigcup C\bigcup \{K\}$ is the node set, and $E$ is the edge set. $\forall u,v\in V$ $(u\neq v)$, if $u$ and $v$ are the two ends of an edge in $E$, $u$ and $v$ should fulfil the following conditions:

\begin{itemize}
  \item If $u\in S$ or $v\in S$, then $u$ and $v$ should meet that $\|u-v\|\leq r$;
  \item if $u\notin S$ and $v\notin S$, then $u$ and $v$ should meet that $\|u-v\|\leq R$,
\end{itemize}
where $\|u-v\|$ denotes the Euclidean distance between $u$ and $v$. $r$ and $R$ ($r\leq R$) are the communication radii
of the SN and RN, respectively. Besides, in this paper a path between $u$ and $v$ is denoted by $p(u,v)$.

\begin{myDef}[DCRNP Problem]\label{def1}
For an undirected graph $G=\{V, E\}$, the DCRNP problem searches for an induced subgraph $G'=\{V',E'\}$ of $G$, where $V'=S\bigcup C'\bigcup\{K\}$, and $C'$ is a subset of $C$ with the minimum cardinality such that the following condition is satisfied:
there exists at least one path, which complies with the delay constraint, between each SN and the sink.
\end{myDef}

Let $P_i=\{p_1(s_i,K),p_2(s_i,K),...,p_{k_i}(s_i,K)\}$ ($1\leq i\leq n$) be the $k_i$ paths connecting the SN $s_i$
and the sink $K$, where $p_j(s_i,K)$ ($1\leq j\leq k_i$) denotes the $j$th path between $s_i$ and $K$. Let
$P=\bigcup\limits_{i=1}^nP_i$ denote the set of all the paths connecting the given SNs and the sink $K$.

CDLs on the induced subgraph $G'$ are selected to deploy RNs, thus, the terms RN and CDL are interchangeable in this paper.
Consequently, a node on $G'$ can be a SN, RN or the sink.
Let $\mathcal{N}(p)$ denote all the nodes on path $p$.
Then, let $\mathbb{C}(p)$ denote RNs on path $p$, and the notation
$\mathcal{C}(p)$ represent the hop count of path $p$. Obviously, we have that $\mathcal{C}(p)=\left|\mathcal{N}(p)\right|-1$.

The DCRNP problem can be formulated as an optimization problem:
\begin{subequations}\label{equ1-1}
\begin{align}
&\text{Minimize}\quad \left|C'\right|，\label{equ1-1a}\\
\begin{split}
&s.t.\quad \forall P_i\in P\ (1\leq i\leq n), \\
&\quad \exists p_j(s_i,K)\in P_i\ (1\leq j\leq |P_i|),\mathcal{C}\left(p_j(s_i,K)\right)\leq \Delta_i，\label{equ1-1b}
\end{split}
\end{align}
\end{subequations}
where $\Delta_i$ is the delay constraint for the $i$th SN. Without loss of generality, in this paper
we assume that the given SNs have the same delay constraint, i.e., $\Delta_i=\Delta$, $i\in \{1,2,...,n\}$.

The DCRNP problem has been proved NP-hard \cite{Bhattacharya10}, therefore, this paper proposes a heuristic algorithm
to approximately solve the DCRNP problem. The path connecting the $i$th SN $s_i$ and the sink and satisfying the delay
constraint $\Delta$ is called a \textbf{feasible path} of $s_i$, and the subgraph satisfying the formulation (\ref{equ1-1b})
is called a \textbf{feasible graph} of the DCRNP problem.

\section{Preliminaries and Heuristic}
Feasible graphs for a DCRNP problem should be connected. Thus, we can find an induced subgraph, which is a
tree rooted at the sink and connecting all the SNs, from each feasible graph. Therefore, in this paper, the proposed
algorithm yields a feasible graph that is a tree for each given DCRNP problem.

Given a tree $T$ and two different nodes of $T$, $u$ and $v$, we denote $p_T(u,v)$ as a path from $u$ to $v$ of $T$. $\forall s\in S$, if $\mathcal{C}(p_T(K,s))\leq\Delta$, we call $T$ a feasible tree for the given DCRNP problem.

\begin{figure}
\begin{center}
\includegraphics[height=5cm]{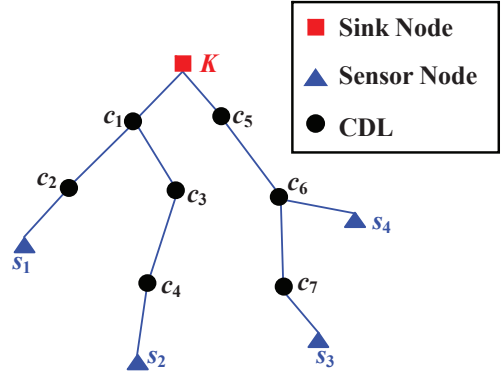}    
\caption{An illustration of the conception of the level.}
\label{fig3}                                 
\end{center}                                 
\end{figure}

Next, the conception of the level of a tree will be introduced to facilitate the explanation of the proposed
SCA algorithm. Given a feasible tree $T$, the level of a node $q$ in $T$ is the hop count of the
path between the sink $K$ and node $q$, i.e., $\mathcal{C}(p_T(K,q))$. The set of nodes on the $k$th level of $T$ is denoted
by $L_T^k$. As illustrated in Fig. \ref{fig3}, $L_T^1=\{c_1,c_5\}$, $L_T^2=\{c_2,c_3,c_6\}$,
$L_T^3=\{s_1,s_4,c_4,c_7\}$ and $L_T^4=\{s_2,s_3\}$.

Observing that the nodes on each level should have
1-hop neighbors on their adjacent levels to maintain the network connectivity, we can deploy RNs
from sink to SNs level by level such that the sink is gradually connected to SNs by the deployed RNs.
Suppose that we have deployed RNs from the 1st level to the $(k-1)$th level of $T$, and now attempt to deploy
RNs for the $k$th level.
Let $\mathcal{V}(L_T^{k-1})$ denote the set of 1-hop neighbors
of the nodes in $L_T^{k-1}$.
Obviously, the nodes on $k$th level should be
1-hop neighbors of nodes on $(k-1)$th level, i.e., $L_T^k\subseteq\mathcal{V}(L_T^{k-1})$.

Therefore, we move on to search the 1-hop neighbors of the nodes on $(k-1)$th level. As the search
is accomplished, i.e., $\mathcal{V}(L_T^{k-1})$ is known, we try to select the nodes on $k$th level
from $\mathcal{V}(L_T^{k-1})$. The nodes that are selected to be located on the $k$th level should
guarantee the existence of feasible paths for the unconnected SNs. For this purpose, an approach, which
represents each node $q$ in $\mathcal{V}(L_T^{k-1})$ by the unconnected SNs that can be connected to the
sink by the feasible paths passing through $q$, is designed and shown in the following path of this section.

In this paper, the shortest path between $u$ and $v$ means a path connecting $u$ and $v$ with least hop count, and the
notation $\mathcal{H}(u,v)$ stands for a shortest path between $u$ and $v$.
Let $\bar{P}_i$ $(1\leq i\leq n)$ be the set of all feasible paths between a SN $s_i$ and the $K$, i.e., $\forall \mathcal{C}(p(s_i,K))\leq\Delta,\ p(s_i,K)\in \bar{P}_i$. Consequently, the set of nodes lying on the feasible paths
of $s_i$ is $\bigcup\limits_{x\in \bar{P}_i}\mathcal{N}\left(x\right)$. Let $q$ be an arbitrary CDL or SN.

\begin{theorem}\label{theorem1}
The sufficient and necessary condition for $q\in\bigcup\limits_{x\in \bar{P}_i}\mathcal{N}\left(x\right)$ is that
\begin{equation}\label{equ3-1}
\mathcal{C}(\mathcal{H}(q,s_i))+\mathcal{C}(\mathcal{H}(q,K))\leq\Delta.
\end{equation}
\end{theorem}

\begin{IEEEproof}
Sufficiency:
From $q\in\bigcup\limits_{x\in P_i}\mathcal{N}\left(x\right)$ we knows that there is at least one path, $p$ ($p\in P_i$),
passing through $q$, therefore, we have $\mathcal{C}(p)\leq\Delta$.

Let $\bar{p}$ ($\tilde{p}$) be a segment of $p$ between $q$ and $s_i$ ($K$). Then, we have
\begin{equation}\label{equ3-3}
\mathcal{C}(\bar{p})\geq \mathcal{C}(\mathcal{H}(q,s_i))
\end{equation}
and
\begin{equation}\label{equ3-4}
\mathcal{C}(\tilde{p})\geq \mathcal{C}(\mathcal{H}(q,K)).
\end{equation}

Since $p$ consists of $\bar{p}$ and $\tilde{p}$, $p$ is a feasible path, which implies
\begin{equation}\label{equ3-5}
\mathcal{C}(p)=\mathcal{C}(\bar{p})+\mathcal{C}(\tilde{p})\leq\Delta.
\end{equation}

Plunging inequalities (\ref{equ3-3}) and (\ref{equ3-4}) into (\ref{equ3-5}), we can achieve
\begin{equation}\label{equ3-6}
\mathcal{C}(\mathcal{H}(q,s_i))+\mathcal{C}(\mathcal{H}(q,K))\leq\Delta.
\end{equation}
So far, we have completed the first part of the proof.

Necessity:
We build two shortest paths which connect $q$ and $s_i$ and connect $q$ and $K$, respectively, i.e.,
$\mathcal{H}(q,s_i)$ and $\mathcal{H}(q,K)$.
Then, we can form a path $p$ between $s_i$ and $K$ by combining $\mathcal{H}(q,s_i)$ and $\mathcal{H}(q,K)$.
According to the inequality (\ref{equ3-1}), the hop count of $p$ is given by
\begin{equation}\label{equ3-9}
\begin{split}
\mathcal{C}(p) &=\mathcal{C}(\mathcal{H}(q,s_i))+\mathcal{C}(\mathcal{H}(q,K))\\
&\leq\Delta,
\end{split}
\end{equation}
which confirms that $p$ is a feasible path between $s_i$ and $K$. Therefore, we can conclude that
\begin{equation}\label{equ3-10}
q\in\bigcup\limits_{x\in \bar{P}_i}\mathcal{N}\left(x\right).
\end{equation}

This completes the proof of Theorem \ref{theorem1}.
\end{IEEEproof}

Theorem \ref{theorem1} implies that if a RN or SN $q$ lies on a feasible path of a SN, then $q$ should satisfy inequality (\ref{equ3-1}).
Let $\bar{S}_k$ be the set of unconnected SNs, which are located on the levels higher than $k$.
As mentioned above, we expect to select the nodes on the $k$th level of $T$ from $\mathcal{V}(L_T^{k-1})$ such that feasible
paths exist for the unconnected SNs. To this end, for each node $q$ in $\mathcal{V}(L_T^{k-1})$,
we try to find the unconnected SNs, which can be connected to sink by the feasible paths passing through
$q$, and the set of these SNs is denoted by $\mathcal{Q}(q)$.
According to Theorem \ref{theorem1}, the SNs in $\mathcal{Q}(q)$ should fulfil that
\begin{equation}\label{equ3-12}
\mathcal{C}(\mathcal{H}(s,q))+k\leq\Delta,
\end{equation}
where $s\in \mathcal{Q}(q)$, and we say that $s$ is connected to the sink by $q$.

When $q$ ($q\in\mathcal{V}(L_T^{k-1})$) is selected as a node on the $k$th level, according to Theorem \ref{theorem1},
we can promise the existence of feasible paths, which pass through $q$ and connect the SNs in $\mathcal{Q}(q)$ to the sink
within delay constraint.

\begin{figure}
\begin{center}
\includegraphics[height=3.5cm]{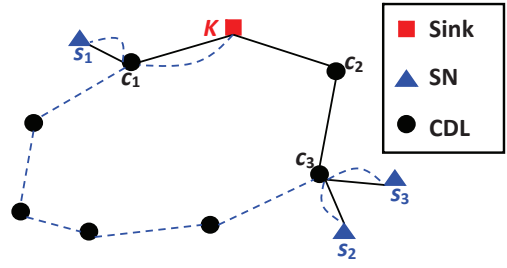}    
\caption{An illustration of the redundant RNs brought by the utilization of inequality (\ref{equ3-12}).}
\label{figp4-1}                                 
\end{center}                                 
\end{figure}

However, the inequality (\ref{equ3-12}) cannot ensure that the RNs deployed on current level are closer to the unconnected SNs than the nodes on previous levels, which may bring in a large amount of redundant deployed RNs. As shown in Fig. \ref{figp4-1}, delay constraint is set as $\Delta=7$. According to inequality (\ref{equ3-12}), we can obtain that $\mathcal{Q}(c_1)=\{s_1,s_2,s_3\}$ and $\mathcal{Q}(c_2)=\{s_2,s_3\}$. Since $\mathcal{V}(L_T^0)=\{c_1,c_2\}$ and all the SNs can be connected to the sink by $c_1$, to select as fewer CDLs as possible on each level, only $c_1$ is selected to be allocated on the 1st level. This ignores that the feasible path that passes through $c_1$ and connects $s_2$ and $s_3$ to the sink is
much longer than the feasible path passing through $c_2$ to $s_2$ and $s_3$, i.e., $\mathcal{C}(p(c_1,s_2))>\mathcal{C}(p(c_2,s_2))$ and $\mathcal{C}(p(c_1,s_3))>\mathcal{C}(p(c_2,s_3))$.
This ignorance leads to a local optimum, whose edges are denoted by the blue dashed line segments in Fig. \ref{figp4-1}, and misses the optimal solution, whose edges are represented by the black line segments. As a result, totally 3 redundant RNs are introduced.

To this end, we introduce a rule to guarantee that the RNs deployed on each level gradually move closer to the unconnected SNs.
Let $\bar{q}$ be a node on the $(k-1)$th level, and $q$ be a 1-hop neighbor of $\bar{q}$.
\begin{lemma}\label{lemma1}
$\mathcal{Q}(q)\subseteq \mathcal{Q}(\bar{q})$.
\end{lemma}
\begin{IEEEproof}
For each SN in $\mathcal{Q}(q)$, the inequality (\ref{equ3-12}) is met, i.e.,
\begin{equation}\label{equ3-13}
\forall s\in\mathcal{Q}(q),\ \mathcal{C}(\mathcal{H}(s,q))+k\leq\Delta.
\end{equation}
Since $q$ is a 1-hop neighbor of $\bar{q}$ and $\bar{q}$ is a node on the $(k-1)$th level, the following
inequality is fulfilled
\begin{equation}\label{equ3-13}
\forall s\in \mathcal{Q}(q),\ \exists p(s,\bar{q}),\ \mathcal{C}(p(s,\bar{q}))+k-1\leq\Delta.
\end{equation}
Due to the fact that $\mathcal{C}(p(s,\bar{q}))\geq\mathcal{C}(\mathcal{H}(s,\bar{q}))$,
we can conclude that
\begin{equation}\label{equ3-113}
\forall s\in \mathcal{Q}(q),\ \mathcal{C}(\mathcal{H}(s,\bar{q}))+k-1\leq\Delta.
\end{equation}
which implies that $\forall s\in \mathcal{Q}(q)$, $s\in \mathcal{Q}(\bar{q})$. Thus $\mathcal{Q}(q)$ is a subset of $\mathcal{Q}(\bar{q})$. This completes the proof.
\end{IEEEproof}

According to Lemma \ref{lemma1}, the SNs in $\mathcal{Q}(q)$ can also be connected to the sink by $\bar{q}$.
Therefore, to guarantee that the nodes on the $k$th level are closer to the unconnected SNs than their 1-hop neighbors on the
$(k-1)$th level, we restrict the nodes on the $k$th level to satisfy a rule
\begin{equation}\label{equ3-14}
\mathcal{C}(\mathcal{H}(q,s))<\mathcal{C}(\mathcal{H}(\bar{q},s)),
\end{equation}
where $s\in\mathcal{Q}(q)$ and $s\in\mathcal{Q}(\bar{q})$. Finally, the SNs in $\mathcal{Q}(q)$ should meet both the inequalities (\ref{equ3-12}) and (\ref{equ3-14}).

\begin{figure}
\begin{center}
\includegraphics[height=5cm]{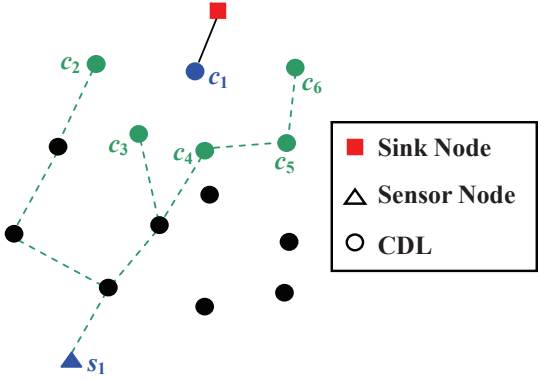}    
\caption{An illustration of how to determine the SNs connected to the sink by a CDL.}
\label{fig5}                                 
\end{center}                                 
\end{figure}

As illustrated in Fig. \ref{fig5}, $c_1$ is the only node on the 1st level, and $s_1$ is connected to the sink by $c_1$.
Additionally, the delay constraint is set as 6, and $\mathcal{C}(\mathcal{H}(c_1,s_1))=4$.
The 1-hop neighbors of $c_1$ are represented by the green circles, i.e., $c_2$, $c_3$, $c_4$, $c_5$ and $c_6$.
The shortest paths from these 1-hop neighbors to $s_1$ are represented by the green dashed line segments.
It can be seen from Fig. \ref{fig5} that the inequality (\ref{equ3-12}) is met by these 1-hop neighbors except
for $c_6$, but only $c_3$ and $c_4$ have a hop count to $s_1$ less than 4. Thus, in these 1-hop neighbors,
only $c_3$ and $c_4$ fulfil both the formulations (\ref{equ3-12}) and (\ref{equ3-14}), i.e., $s_1$ can be
connected to the sink only by $c_3$ and $c_4$.

To connect all the unconnected SNs to the sink, we should select a subset $U$ of $\mathcal{V}(L_T^{k-1})$ such
that each unconnected SN can be connected to the sink by these nodes, i.e., $\bigcup\limits_{u\in U}\mathcal{Q}(u)=\bar{S}_k$.
Furthermore, to save the deployed RNs we expect to minimize the cardinality of subset $U$.
Thus, we can formulate the problem, which requires to select the nodes on $k$th level from $\mathcal{V}(L_T^{k-1})$,
as the set covering problem described follows: given an unconnected SNs set $\bar{S}_k$ and a node set $\mathcal{V}(L_T^{k-1})$,
whose each element $q$ is represented by the unconnected SNs that can be connected to the sink by $q$, then this problem searches
for a minimum subset $U$ of $\mathcal{V}(L_T^{k-1})$ such that $\bigcup\limits_{u\in U}\mathcal{Q}(u)=\bar{S}_k$.
As a result, we can employ the set covering algorithm to solve this problem. Due to the effective
performance and the wide utilization, the greedy heuristic algorithm is employed to solve this set covering problem.

\section{Description of SCA}
The proposed SCA algorithm is composed of three steps and detailed in Algorithm \ref{SCA}. In the first step, the SCA algorithm preliminarily tests
either a feasible solution exists or RNs are necessary for the DCRNP problem based on shortest path tree algorithm.

\begin{algorithm}[htb]         
\footnotesize
\caption{SCA.}             

\label{SCA}                  

\begin{algorithmic}[1]                

\REQUIRE ~~\\                          
   A set of SNs $S=\{s_1,s_2,...,s_n\}$, a set of CDLs $C=\{c_1,c_2,...,c_m\}$, a sink node $K$,
   the communication radii of the SN $r$ and the RN $R$, the delay constraint $\Delta$.

\ENSURE ~~\\                           
    A feasible tree $T$.
\STATE $tmp=$a shortest path tree rooted at $K$ and including all SNs in $S$ with using CDLs in $C$; \%the first step of SCA begins.
\IF{$\forall i\in\{1,2,...,n\},\ p_{tmp}(s_i,K)\leq\Delta$}
    \STATE $tmp=$a shortest path tree rooted at $K$ and including all SNs in $S$ without any CDLs in $C$;
    \IF{$\exists i\in\{1,2,...,n\},\ p_{tmp}(s_i,K)>\Delta$}
        \STATE input $S$, $C$ and $K$ into the second step of SCA;   \%the first step of SCA ends.
        \STATE $C'=$a subset of $C$ returned by the second step of SCA;
        \STATE input $S$, $C'$ and $K$ into the third step of SCA;
        \STATE $T=$a feasible tree returned by the third step of SCA;
    \ELSE
        \STATE declare that no RNs are necessary to build a feasible tree $T$ and terminate SCA;
    \ENDIF
\ELSE
    \STATE declare that no feasible solutions exist for this DCRNP problem and terminate SCA;
\ENDIF
\RETURN $T$;

\end{algorithmic}

\end{algorithm}

The second step is the principal body of the SCA algorithm. The second step of SCA is iteratively
executed so as to allocate the nodes on each level. To be specific, in the $k$th iteration, SSCA moves on
to select nodes on the $k$th level. Firstly, the second step of SCA searches for 1-hop neighbors of the nodes on the $(k-1)$th level,
i.e., $\mathcal{V}(L_T^{k-1})$. Next, for each node $q$ in $\mathcal{V}(L_T^{k-1})$, the second step of SCA searches for the
unconnected SNs, which can be connected to the sink by $q$, i.e., $\mathcal{Q}(q)$. Then, each node $q$ in
$\mathcal{V}(L_T^{k-1})$ is represented by $\mathcal{Q}(q)$, and as a result the current problem is transformed into
the set covering problem, which attempts to select a minimum subset of $\mathcal{V}(L_T^{k-1})$ to fully cover the unconnected SNs.
Finally, the greedy heuristic algorithm is employed to solve this set covering problem. Consequently, the selected
subset of $\mathcal{V}(L_T^{k-1})$ is the set of nodes on $k$th level. As the execution of the second step of SCA, a feasible
tree $T$ is gradually built, and clearly, due to the delay constraint, the number of iteration of the second step of SCA cannot
be larger than $\Delta$.

\begin{figure}
\begin{center}
\includegraphics[height=5cm]{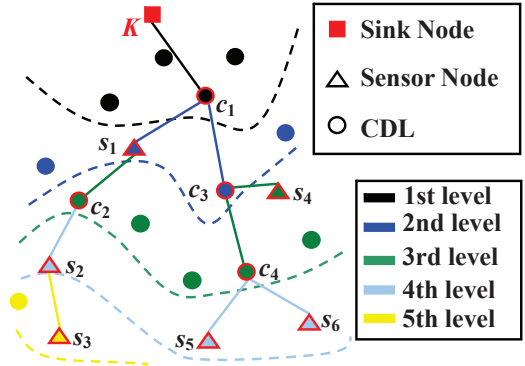}    
\caption{An illustration of the procedure of the second step of SCA.}
\label{fig4}                                 
\end{center}                                 
\end{figure}

The process of the second step of SCA is illustrated in Fig. \ref{fig4}, where the circles and triangles with
the same color are the 1-hop neighbors of the nodes on the same level, and the circles and triangles
with red edges are the SNs or RNs on each level. At the beginning, the second step of SCA
first searches the 1-hop neighbors of the sink $K$, and these neighbors are denoted by the black
circles and triangles. Next, the second step of SCA represents these neighbors by the unconnected SNs,
and the greedy heuristic is employed to find a set cover for
the unconnected SNs. As shown in Fig. \ref{fig4}, $c_1$ is selected in this iteration. In the
second iteration, the second step of SCA first finds the
1-hop neighbors of $c_1$, and these neighbors are denoted by the blue circles and triangles. Then,
$s_1$ and $c_3$ are selected in this iteration. As the iteration repeats, all the unconnected
SNs are gradually connected to the sink. The second step of SCA
is detailed in the Algorithm \ref{SSCA}.

\begin{remark}\label{remark1}
As 1-hop neighbors have been represented by the SNs connected to the sink by them, the situation
that several neighbors connect the same unconnected SNs to the sink may happen, i.e.,
$\mathcal{Q}(q)=\mathcal{Q}(\bar{q})$, where $q$ and $\bar{q}$ are all 1-hop neighbors of the nodes on the
last level. To deal with this
situation, the weight of a neighbor, $q$, is introduced and defined as follows. Let $\mathcal{Q}(q)$ be the
set of unconnected SNs, which are connected to the sink by $q$. Then the weight, $\omega(q)$, of
$q$ is given by
\begin{equation}\label{equ3-15}
\omega(q)=\left|\bigcup\limits_{x\in \mathcal{Q}(q)}\mathbb{C}(\mathcal{H}(q,x))\right|.
\end{equation}
If the situation mentioned above happens, we select the one with least weight.
\end{remark}

\begin{algorithm}[htb]         
\footnotesize
\caption{The Second step of SCA.}             

\label{SSCA}                  

\begin{algorithmic}[1]                

\REQUIRE ~~\\                          
   A set of SNs $S=\{s_1,s_2,...,s_n\}$, a set of CDLs $C=\{c_1,c_2,...,c_m\}$, a sink node $K$,
   the communication radii of the SN $r$ and the RN $R$, the delay constraint $\Delta$.

\ENSURE ~~\\                           
    A set, $C'$, of CDLs.

\STATE $\mathcal{V}(L_T^0)$ = the 1-hop neighbors of $K$ in $S$ and $C$;
\STATE $L_T^0=K$;
\STATE $k=1$;
\STATE $\bar{S}_k$ = $S$;
\WHILE{($\bar{S}\neq\varnothing$)}
    \STATE $\bar{S}_k=\bar{S}_k-\mathcal{V}(L_T^0)$;
    \STATE $i=0$;
    \STATE $tmp =$the $i$th element in $\mathcal{V}(L_T^0)$;
    \WHILE{($tmp\neq \varnothing$)}
        \STATE $con[i]=\mathcal{Q}(tmp)$;
        \STATE $weight[i]= \omega(tmp)$;
        \STATE $i=i+1$;
        \STATE $tmp=$the $i$th element in $\mathcal{V}(L_T^0)$;
    \ENDWHILE
    \STATE sort $\mathcal{V}(L_T^0)$ according to the cardinality of each $con[i]$ and the $weight[i]$;
    \STATE $L_T^k$ = a subset of $\mathcal{V}(L_T^0)$ found by the greedy heuristic algorithm;
    \STATE $\bar{S}_k=\bar{S}_k-L_T^k$;
    \STATE $C=C-L_T^k$;
    \STATE $\mathcal{V}(L_T^k)$ = the 1-hop neighbors, which are selected from $\bar{S}_k$ and $C$, of the nodes in $L_T^k$;
    \STATE $tmpRe=L_T^k-S$;
    \STATE $C'=C'\bigcup tmpRe$;
    \STATE $k=k+1$;
\ENDWHILE
\RETURN $C'$;

\end{algorithmic}

\end{algorithm}

\begin{figure}
\begin{center}
\includegraphics[height=4cm]{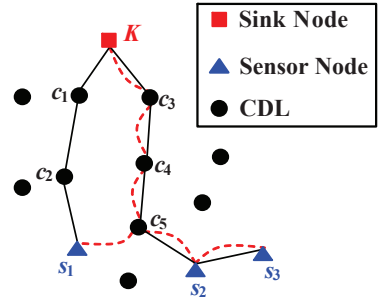}    
\caption{An illustration of the redundant CDLs introduced by the second step SCA.}
\label{fig6}                                 
\end{center}                                 
\end{figure}

The redundant RNs may be deployed by the second step SCA for the introduction of the rule (\ref{equ3-14}),
which is illustrated in Fig. \ref{fig6}.
The feasible tree returned by the second step SCA is denoted by the black line segments in Fig. \ref{fig6}.
Since $\mathcal{C}(\mathcal{H}(K, s_1))<\mathcal{C}(\mathcal{H}(c_3, s_1))$,
$s_1$ cannot be connected to the sink
by $c_3$, and the second step SCA select two CDLs to deploy RNs, i.e., $c_1$ and $c_3$, on the first level.
However, there is a feasible path passing through $c_3$ and connecting $K$ to $s_1$. Thus,
the redundant RNs $c_1$ and $c_2$ are introduced. Actually, a solution
with fewer RNs exists in Fig. \ref{fig6}, and this solution is represented by the red
dashed line segments.

To save the redundant RNs introduced by the second step of SCA, the third step of the SCA algorithm is designed. Since we expect that the number of RNs deployed on each level is as fewer
as possible, we try remove the RNs connecting fewer SNs to the sink on each level.
Thus, each RN $q$ returned by the second step of SCA is assigned a weight, which is defined
as the number of SNs meeting the inequality (\ref{equ3-1}) for $q$. Then, the returned RNs
are sorted in ascending order according to the weight, and we try to gradually delete the redundant RNs
from the one with least weight. As a RN is deleted, the shortest path tree is formed to check
whether a feasible tree exists for the DCRNP problem on the remaining RNs. If such feasible
tree exists, the RN is deleted forever, otherwise the RN is retrieved and marked as tried.
This procedure is repeated until all the RNs is deleted or
marked as tried. The third step of the SCA algorithm is detailed in the Algorithm \ref{TSCA}.

\begin{algorithm}[htb]         
\footnotesize
\caption{The Third step of SCA.}             

\label{TSCA}                  

\begin{algorithmic}[1]                

\REQUIRE ~~\\                          
   The set $S$ of SNs, a set $C'$ of CDLs returned by the second step of SCA.

\ENSURE ~~\\                           
    A feasible tree $T$.

\STATE $T=\text{a}$ shortest path tree rooted at the sink and including all the
given SNs by using the CDLs in $C'$;
\FORALL{$tmp\in C'$}
    \STATE calculate the weight of $tmp$;
\ENDFOR
\STATE sort the CDLs in $C'$ in an ascending order according the weight;
\STATE $tmp=\text{the}$ first element of $C'$;
\STATE $C'=C'-tmp$;
\WHILE{($tmp\neq\varnothing$)}
    \STATE $tmpT=\text{a}$ shortest path tree rooted at the sink and including all the
    given SNs by using the CDLs in $C'$;
    \IF{($tmpT$ is a feasible tree)}
        \STATE $T$ = $tmpT$;
        \STATE $C'=\text{the}$ CDLs on $T$;
        \STATE sort the CDLs in $C'$ in an ascending order according the weight;
        \STATE $tmp=\text{the}$ first element of $C'$;
        \STATE $C'=C'-tmp$;
    \ELSE
        \STATE mark $tmp$ as tried;
        \STATE $C'=C'\bigcup tmp$;
        \STATE $tmp=\text{the}$ least weighted untried CDL in $C'$;
    \ENDIF
\ENDWHILE
\RETURN $T$;

\end{algorithmic}

\end{algorithm}

\section{Analysis of SCA}
\subsection{Approximation Ratio}
Let $T^*$ be a optimal feasible tree for the given DCRNP problem, and the set of the RNs and SNs on the $k$th level
of $T^*$ is denoted by $L^k_{T^*}$. Let $T$ be a feasible tree returned by the SRNP algorithm,
and the set of RNs and SNs on the $k$th level of $T$ is denoted by $L_{T^*}^k$. Thus, the ratio between the optimal solution
and the solution returned by the SCA algorithm is given by
\begin{equation}\label{equ4-1}
\frac{\left|\bigcup\limits_{k=1}^lL_T^k-S\right|}{\left|\bigcup\limits_{k=1}^{l^*}L^k_{T^*}-S\right|}
=\frac{\left|\bigcup\limits_{k=1}^l\left(L_T^k-S\right)\right|}{\left|\bigcup\limits_{k=1}^{l^*}\left(L^k_{T^*}-S\right)\right|},
\end{equation}
where $l$ and $l^*$ denote the number of levels of $T$ and $T^*$, respectively.

Because of the fact that a node can only be located on a certain level, i.e., $L_T^i\bigcap L_T^j=\varnothing$ and
$L_{T^*}^i\bigcap L_{T^*}^j=\varnothing$  $(i\neq j)$, equality (\ref{equ4-1}) changes into
\begin{equation}\label{equp4-2}
\begin{split}
\frac{\left|\bigcup\limits_{k=1}^l\left(L_T^k-S\right)\right|}{\left|\bigcup\limits_{k=1}^{l^*}\left(L^k_{T^*}-S\right)\right|}
&=\frac{\sum\limits_{k=1}^l\left|L_T^k-S\right|}{\sum\limits_{k=1}^{l^*}\left|L_{T^*}^k-S\right|}\\
&=\frac{\sum\limits_{k=1}^l\left|L_T^k\right|-|S|}{\sum\limits_{k=1}^{l^*}\left|L_{T^*}^k\right|-|S|},
\end{split}
\end{equation}
which leads to that
\begin{equation}\label{equp4-3}
\begin{split}
\frac{\sum\limits_{k=1}^l\left|L_T^k\right|-|S|}{\sum\limits_{k=1}^{l^*}\left|L_{T^*}^k\right|-|S|}&<
\frac{\sum\limits_{k=1}^l\left|L_T^k\right|}{\sum\limits_{k=1}^{l^*}\left|L_{T^*}^k\right|}\\
&=\frac{|L_T^1|}{\sum\limits_{k=1}^{l^*}\left|L_{T^*}^k\right|}+\frac{|L_T^2|}{\sum\limits_{k=1}^{l^*}\left|L_{T^*}^k\right|}+
...+\frac{|L_T^l|}{\sum\limits_{k=1}^{l^*}\left|L_{T^*}^k\right|}.
\end{split}
\end{equation}

Let $OPT_k$ be a minimum set cover for the $k$th level of $T$, then, we can obtain that
\begin{equation}\label{equp4-4}
\footnotesize
\begin{split}
\frac{|L_T^1|}{\sum\limits_{k=1}^{l^*}\left|L_{T^*}^k\right|}+\frac{|L_T^2|}{\sum\limits_{k=1}^{l^*}\left|L_{T^*}^k\right|}+
...+\frac{|L_T^l|}{\sum\limits_{k=1}^{l^*}\left|L_{T^*}^k\right|}=\\
\frac{|L_T^1|}{|OPT_1|}\frac{|OPT_1|}{\sum\limits_{i=1}^{l^*}\left|L_{T^*}^k\right|}+\frac{|L_T^2|}{|OPT_2|}\frac{|OPT_2|}{\sum\limits_{k=1}^{l^*}\left|L_{T^*}^k\right|}+
...&+\frac{|L_T^l|}{|OPT_l|}\frac{|OPT_l|}{\sum\limits_{k=1}^{l^*}\left|L_{T^*}^k\right|}.
\end{split}
\end{equation}

Since greedy algorithm is employed to solve the set covering problem that requires to select a subset of $V(L_T^{k-1})$ to cover
the unconnected SNs $\bar{S}_k$ on the $k$th level, considering that the approximation ratio of greedy set covering algorithm is
$\ln|\bar{S}_k|$ \cite{Cormen01}, we have that
\begin{equation}\label{equp4-5}
\forall k\in\{1,2,...,l\},\ \frac{|L_T^k|}{|OPT_k|}\leq \ln|\bar{S}_k|.
\end{equation}

In the first step of SCA, we check the existence of the feasible solution without RNs for DCRNP problem,
thus, for the DCRNP problem input to the second step, its optimal solution contains at least one RN, i.e.,
\begin{equation}\label{equp4-06}
\left|\bigcup\limits_{k=1}^{l^*}L^k_{T^*}-S\right|\geq 1,
\end{equation}
and according to equations (\ref{equ4-1})-(\ref{equp4-2}) this leads to that
\begin{equation}\label{equp4-6}
\sum\limits_{k=1}^{l^*}\left|L_{T^*}^k\right|\geq |S|+1.
\end{equation}

Besides, due to the fact that $|OPT_k|\leq |\bar{S}_k|\leq |S|$, combining with inequalities (\ref{equp4-5})-(\ref{equp4-6}),
the formulation (\ref{equp4-4}) changes into
\begin{equation}\label{equp4-7}
\footnotesize
\begin{split}
\frac{|L_T^1|}{|OPT_1|}\frac{|OPT_1|}{\sum\limits_{k=1}^{l^*}\left|L_{T^*}^k\right|}+&\frac{|L_T^2|}{|OPT_2|}\frac{|OPT_2|}{\sum\limits_{k=1}^{l^*}\left|L_{T^*}^k\right|}+
...+\frac{|L_T^l|}{|OPT_l|}\frac{|OPT_l|}{\sum\limits_{k=1}^{l^*}\left|L_{T^*}^k\right|}\\
&\leq l\ln|S|\frac{|S|}{|S|+1}\\
&\leq \Delta\ln|S|.
\end{split}
\end{equation}

Finally, the approximation ratio of SCA can be represented by
\begin{equation}\label{equp4-8}
\begin{split}
\frac{\left|\bigcup\limits_{k=1}^lL_T^k-S\right|}{\left|\bigcup\limits_{k=1}^{l^*}L^k_{T^*}-S\right|}=\text{O}(\ln|S|).
\end{split}
\end{equation}

\subsection{Time Complexity}
Let $N=|S|+|C|+1$, then the time
complexity of the shortest path tree algorithm is given by $N\lg N$ \cite{Cormen01}.
The time complexity of the first step of the SRNP algorithm can be easily calculated, i.e.,
$\text{O}(N\lg N)$.

Then we analyze the time complexity of the Algorithm \ref{SSCA}. In each iteration of the second step SCA, the
main loop between lines 5-23 is executed. Therefore, we first analyze the time complexity to execute
the main loop for once.
In the inner loop between lines 9-14, the shortest path tree algorithm is applied to find the SNs
fulfilling the inequalities (\ref{equ3-12}) and (\ref{equ3-14}). Thus, the time complexity of one iteration
of this loop is $\text{O}(N\lg N)$. Additionally, this loop can be iterated at most $N$ times, which leads
to that the time complexity of this loop is $\text{O}(N^2\lg N)$. Then, the greedy heuristic algorithm with a time
complexity of $\text{O}(N^2)$ \cite{Cormen01} is employed to find a minimum set cover for each level. Moreover,
in each iteration of the second step of SCA, we will find the 1-hop neighbors within $\text{O}(N^2)$ running time. Hence,
the time complexity of one iteration of the main loop is
\begin{equation}\label{equ4-7}
\text{O}(N^2\lg N)+\text{O}(N^2)=\text{O}(N^2\lg N).
\end{equation}
Since the main loop will be iterated at most $\Delta$ times, the time complexity of the second step of SCA is $\text{O}(N^2\lg N)$.

In the main loop of the third step of SCA, the shortest path tree algorithm is implemented at most $N$ times, and the time
complexity to sort the remaining CDLs is $\text{O}(N^2)$. Thereby, the time complexity of the third step of SCA is $\text{O}(N^3)$.
Since the SCA algorithm is the combination of the three steps, the total time complexity, $T_{SCA}$, of
the SCA algorithm is given by
\begin{equation}\label{equ4-8}
\begin{split}
T_{SCA}&=\text{O}(N\lg N)+\text{O}(N^2\lg N)+\text{O}(N^3)\\
&=\text{O}(N^3),
\end{split}
\end{equation}
which indicates that the proposed SCA is a polynomial time algorithm.

\section{Simulation Results}
In the simulations, SNs are randomly placed on a square field with side length
of 100 units. To ensure a high probability that the DCRNP problem has a feasible solution, the
amount of CDLs is relative large, i.e., is set as 400, and these CDLs are randomly distributed on
the deployment field. During the
simulation, the amount of SNs is varying from 10 to 100. The simulation is carried out under
both the homogeneous network, i.e., $r=R=10$, $r=R=15$ and $r=R=20$, and the heterogeneous network, i.e.,
$r=10$ and $R=15$. SPTiRP algorithm proposed in \cite{Bhattacharya10}-\cite{Bhattacharya14}
is taken as the baseline. Since the heterogeneous network is not considered in \cite{Bhattacharya10}-\cite{Bhattacharya14},
when simulation is performed under the heterogeneous network, SPTiRP is adapted such that it can be implemented to the heterogeneous network.
The notations used in the remaining part of this section are listed as follows:
\begin{itemize}
  \item $RN(\text{SPTiRP})$ and $RN(\text{SCA})$ represent the number of RNs deployed by SPTiRP and SCA, respectively.
  \item $RN(\text{SPTiRP})-RN(\text{SCA})$ represents the difference of deployed RNs between SPTiRP and SCA.
\end{itemize}

\subsection{Statistic Analysis}

\begin{figure*}[!t]
\centering
\subfigure[$r=10,R=15$.] {\label{fig6-3a}\includegraphics[height=1.5in,width=1.7in,angle=0]{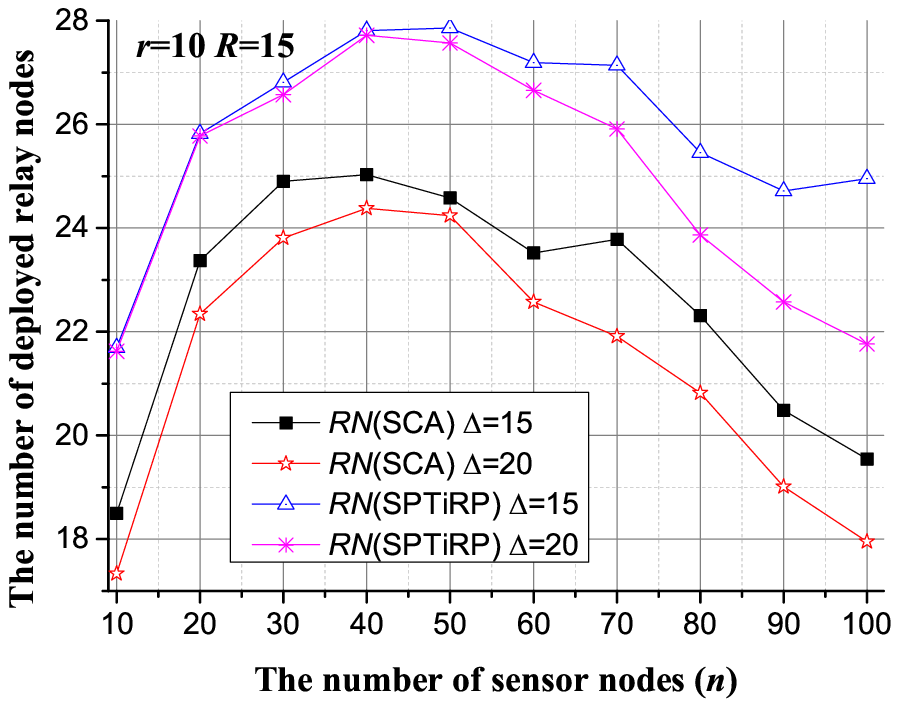}}
\subfigure[$r=R=20$.] {\label{fig6-3b}\includegraphics[height=1.5in,width=1.7in,angle=0]{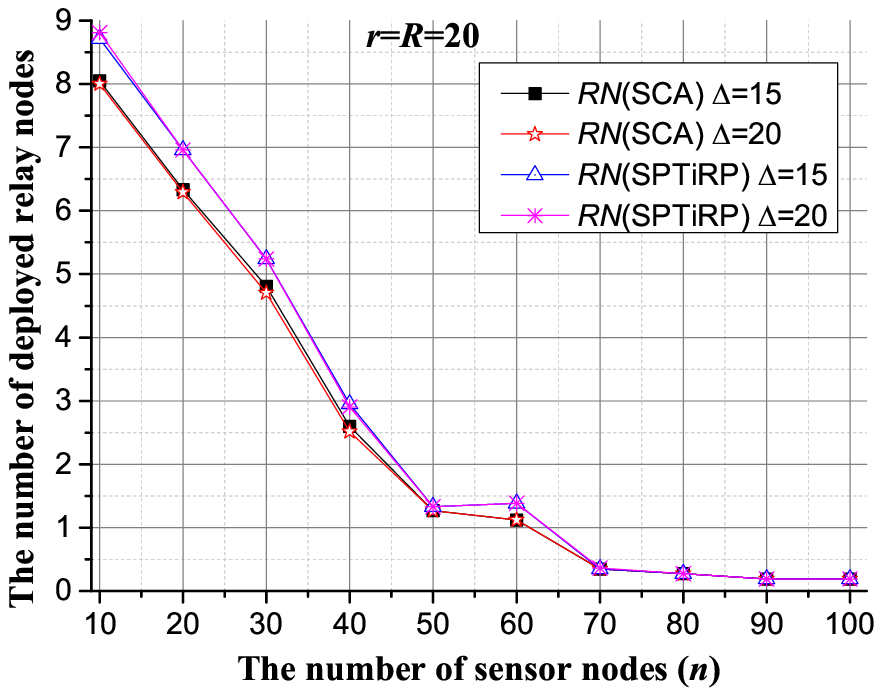}}
\subfigure[$r=R=15$.] {\label{fig6-3c}\includegraphics[height=1.5in,width=1.7in,angle=0]{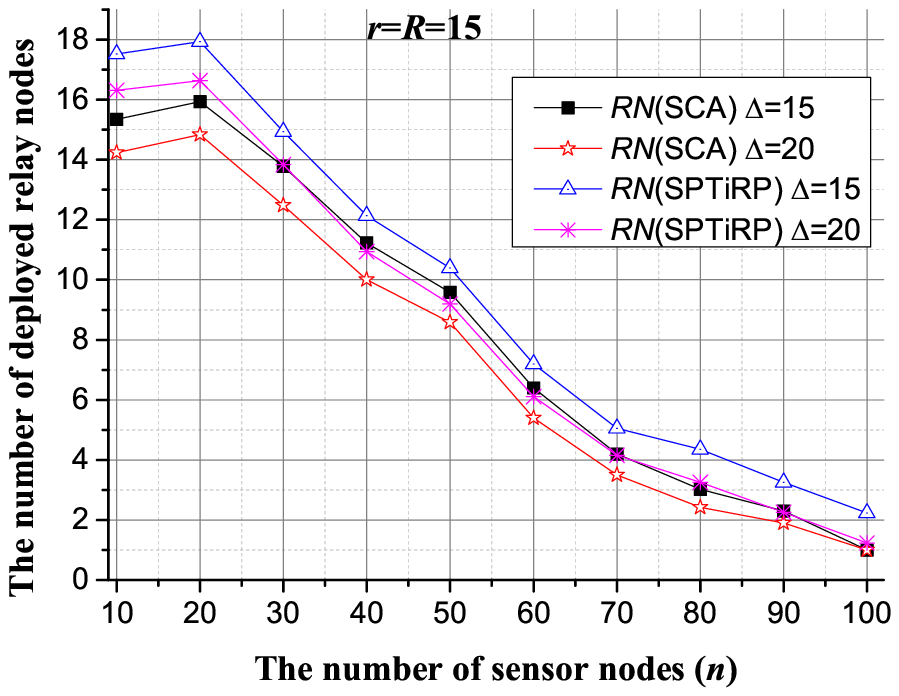}}
\subfigure[$r=R=10$.] {\label{fig6-3d}\includegraphics[height=1.5in,width=1.7in,angle=0]{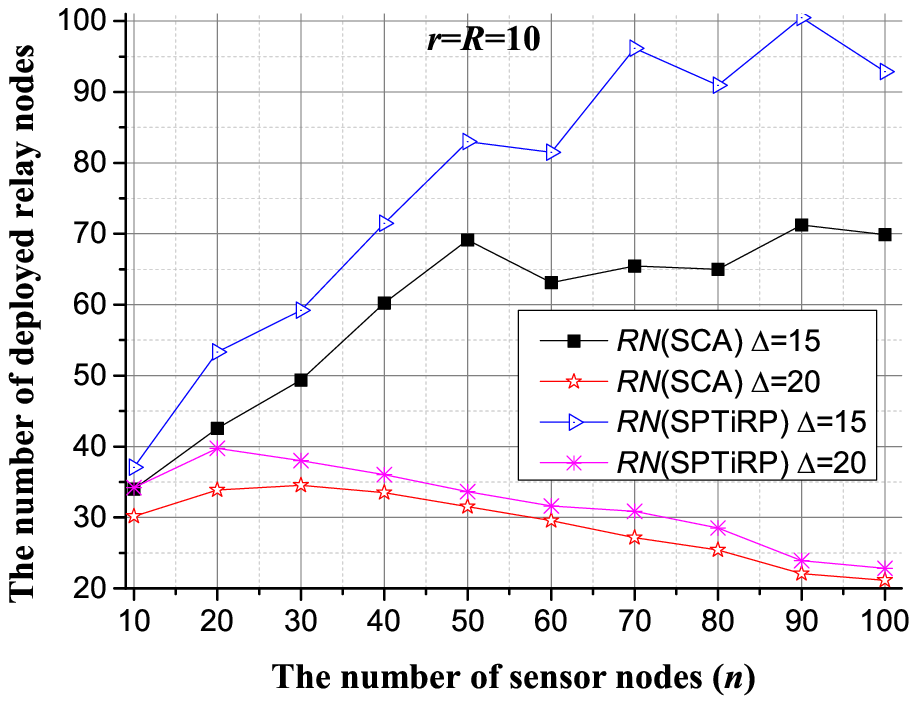}}
\caption{The number of RNs deployed by SPTiRP and SCA under different $r$, $R$ and $\Delta$s.}
\label{fig6-3}
\end{figure*}

\begin{figure*}[!t]
\centering
\subfigure[$r=10,R=15$.] {\label{fig6-4a}\includegraphics[height=1.5in,width=1.7in,angle=0]{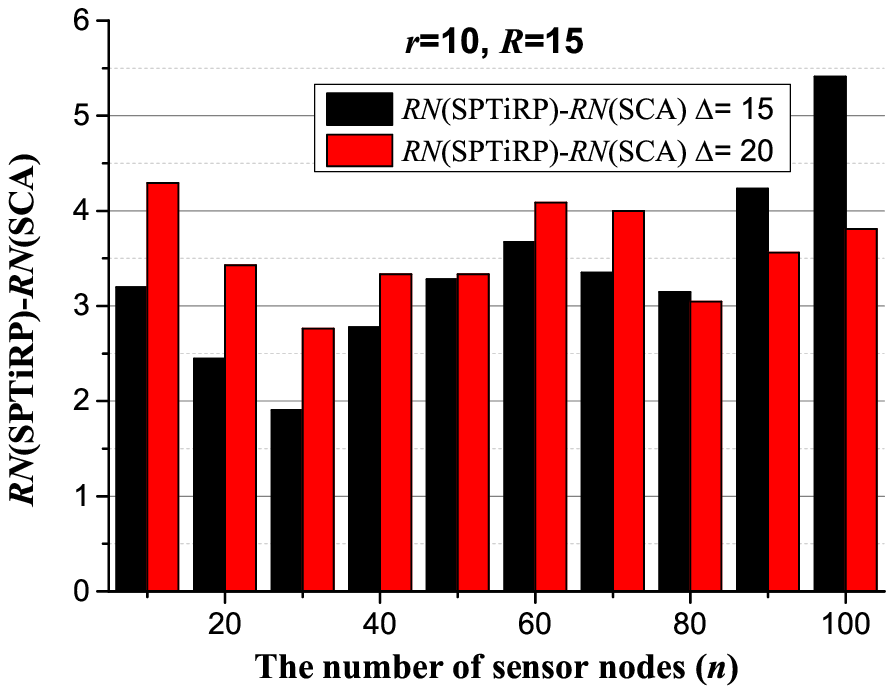}}
\subfigure[$r=R=20$.] {\label{fig6-4b}\includegraphics[height=1.5in,width=1.7in,angle=0]{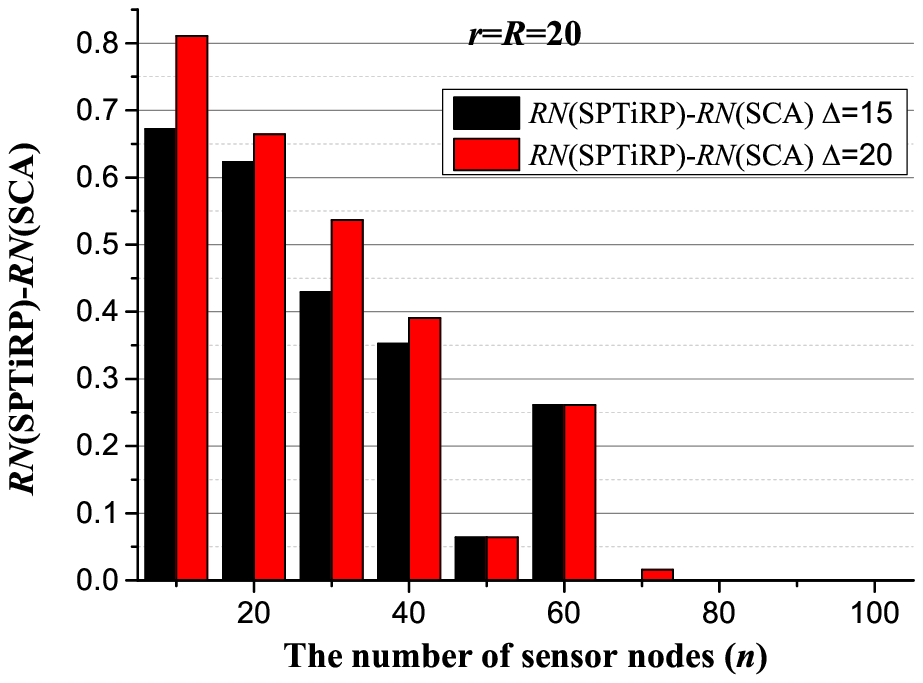}}
\subfigure[$r=R=15$.] {\label{fig6-4c}\includegraphics[height=1.5in,width=1.7in,angle=0]{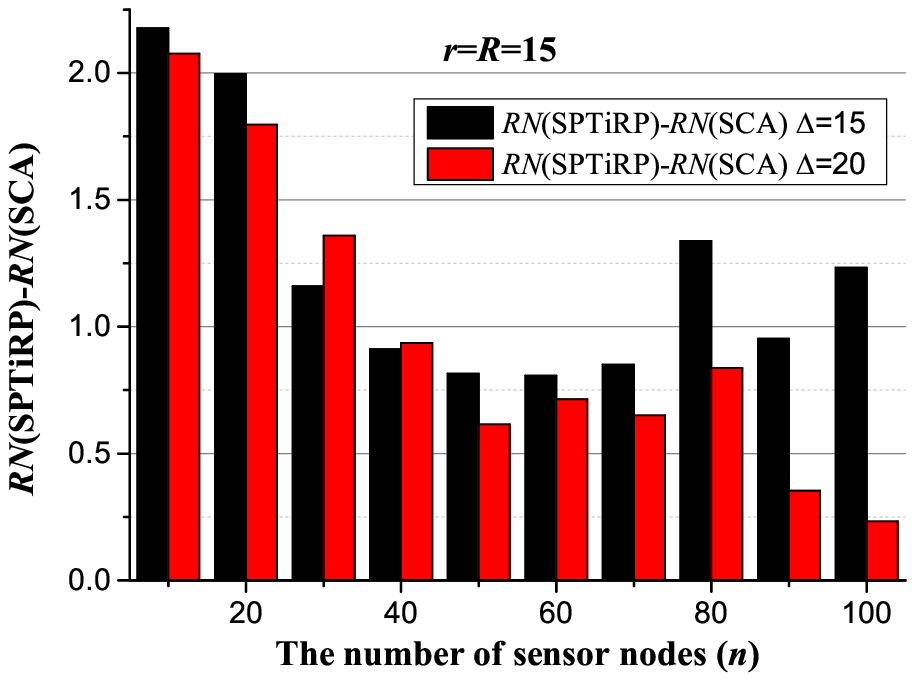}}
\subfigure[$r=R=10$.] {\label{fig6-4d}\includegraphics[height=1.5in,width=1.7in,angle=0]{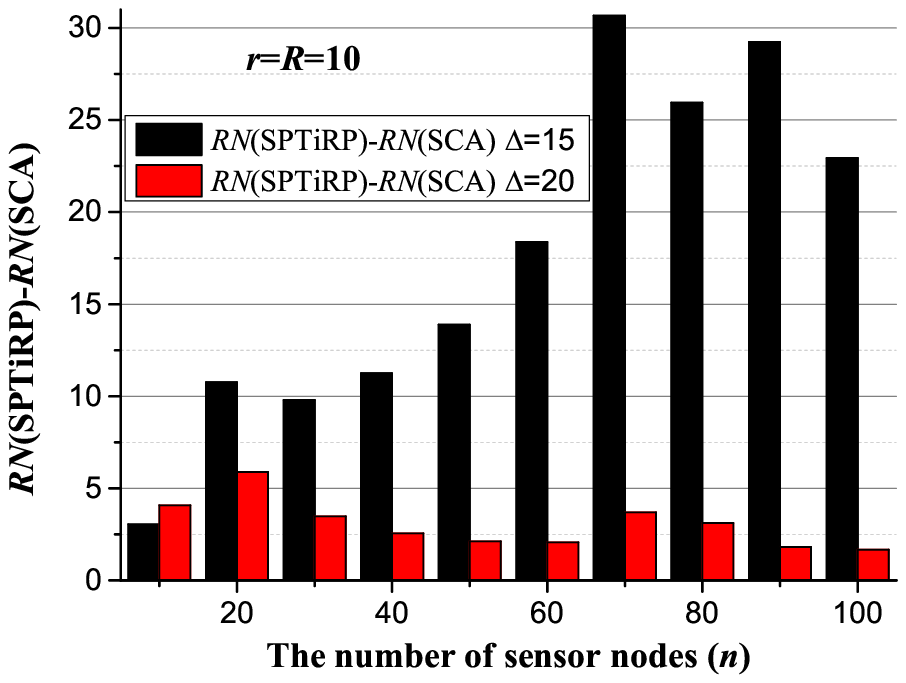}}
\caption{The different of deployed RNs between SPTiRP and SCA under different $r$, $R$ and $\Delta$s.}
\label{fig6-4}
\end{figure*}

The number of RNs deployed by SPTiRP and SCA under different conditions is shown in Fig.
\ref{fig6-3}, and the difference of deployed RNs between SPTiRP and SCA (i.e., $RN(\text{SPTiRP})-RN(\text{SCA})$) under
different conditions is shown in Fig. \ref{fig6-4}.

The results show that RNs deployed by SCA are fewer than those deployed by
SPTiRP under different communication radii and delay constraints, which demonstrates the
efficiency of SCA. It can be seen from Fig. \ref{fig6-3a} that when the communication radii
of SNs and RNs is set as $r=10$ and $R=15$, SCA can save at most 5.4095 $(5.4095/24.9524
\approx 21.7\%)$ deployed RNs comparing to SPTiRP. In Fig. \ref{fig6-3b}, the communication
radii of SNs and RNs is set as $r=R=20$, SCA can save at most 0.8108 $(0.8108/8.8108
\approx 10\%)$ deployed RNs comparing to SPTiRP. In Fig. \ref{fig6-3c}, the communication radii of SNs and RNs is set as $r=R=15$,
and the maximum saving due to SCA is 2.18 ($2.18/17.5\approx 12.5\%$).
In Fig. \ref{fig6-3d}, the communication radii
of SNs and RNs is set as $r=R=10$, SCA can save at most 30.6667
$(30.267/96.1429\approx 31.48\%)$ deployed RNs comparing to SPTiRP.

It is clear in Fig. \ref{fig6-3} that the number of RNs deployed by SPTiRP and SCA both increases with the decreasing
of communication radii of SNs and RNs, and it is easy to explain this by that more RNs are required to connected the whole
network as their communication radii become smaller.
Fig. \ref{fig6-3} also indicate that as the delay constraint is relaxed, SPTiRP
and SCA will deployed fewer RNs under the same communication radii. This is due to
that as the delay constraint is relaxed, the algorithm will select the solution containing fewer RNs,
even though this will yield feasible paths with larger hop count for the given SNs.

The number of RNs saved by SCA comparing to SPTiRP under different conditions is
shown in Fig. \ref{fig6-4}. It can be learned from these figures that as the communication radii becomes smaller,
more RNs can be saved by SCA comparing to SPTiRP, i.e., the value of $RN(\text{SPTiRP})-RN(\text{SCA})$ grows.

This phenomenon can be
explained by that as the communication radii becomes smaller, the topology becomes sparser and the hop
count between the nodes on different paths becomes larger, therefore, the opportunity to merge different paths
only by deleting the deployed RNs becomes smaller. Hence, the performance of SPTiRP deteriorates as
the communication radii becomes smaller. On the other hand, SCA do not confront the limitation suffering by
SPTiRP, therefore, the SCA algorithm will not suffer this deterioration. Consequently, the
difference between these two algorithms grows larger as the communication radii becomes smaller.

Additionally, we can see from Fig. \ref{fig6-4} that as the delay constraint is tighten SCA can save more RNs
than SPTiRP. This is due to that SCA can deploy fewer RNs on each level based on the set covering approach
even when delay constraint becomes smaller. In contrast, when delay constraint is tighten, the RNs can be deleted
from the original shortest path tree by SPTiRP become fewer, which reduce the performance of SPTiRP. As a result,
the difference of deployed RNs between SPTiRP and SCA becomes larger.

\section{Conclusion}
In this paper, the DCRNP problem is studied. For the NP-hard nature of the DCRNP problem,
an approximation algorithm-SCA is proposed to handle this problem based on the set
covering method. The novelty of the proposed SCA algorithm lies in its multi iteration
procedure, and in each iteration the set covering method is applied to select the SNs or
CDLs. The approximation ratio of the SCA algorithm is proved to be $\text{O}(\ln|S|)$.
The time complexity of the SCA algorithm is also rigorously analyzed, and it shows that
the SCA is a polynomial time algorithm with the time complexity of
$\text{O}(N^3)$. Finally, extensive simulations are carried out to
evaluate the performance of the SCA algorithm. The simulation results show that the
SCA algorithm can significantly save the deployed relay nodes comparing to the existing
algorithm within a worthy running time.


%

\appendices


\section*{Acknowledgment}
This work was supported by the Natural Science Foundation of China (61174026, 61233007, and 61304263)
and Cross-disciplinary Collaborative Teams Program for Science, Technology and Innovation of Chinese
Academy of Sciences (Network and System Technologies for Safety Monitoring and Information Interacting
in Smart Grid).

\ifCLASSOPTIONcaptionsoff
  \newpage
\fi

\end{document}